# Quantum Phase Space Symmetry and Sterile Neutrinos

**Ravo Tokiniaina Ranaivoson[1], Raoelina Andriambololona[2], Hanitriarivo Rakotoson[3], Roland Raboanary[4], Joël Rajaobelison[5], Philippe Manjakasoa Randriantsoa[6]**

*tokhiniaina@gmail.com[1], tokiniainaravor13@gmail.com[1],*
*raoelina.andriambololona@gmail.com[2], infotsara@gmail.com[3], r_raboanary@yahoo.fr[4],*
*rajaobelisonjoel@gmail.com[5], njakarandriantsoa@gmail.com[6]*

[1,2,3,5,6]*Information Technology and Theoretical Physics Department*
*Institut National des Sciences et Techniques Nucléaires (INSTN- Madagascar)*
BP 3907 Antananarivo 101, Madagascar, *instn@moov.mg*

[1,2,3] TWAS *Madagascar Chapter, Malagasy Academy*
BP 4279 Antananarivo 101, Madagascar

[3,4,6]*Faculty of Sciences , iHEPMAD – University of Antananarivo,*
BP 566 Antananarivo 101, Madagascar

*Abstract:* On one hand, the concept of Quantum Phase Space which is compatible with the uncertainty principle has been considered recently. It has also been shown that a natural symmetry that can be associated with this quantum phase space is the symmetry corresponding to Linear Canonical Transformations (LCTs). On the other hand, sterile neutrinos are hypothetical particles that are expected to be important for the understanding of physics beyond the current standard model of particle physics. The existence of these particles are suggested both from theoretical and experimental sides. In this work, the objective is to discuss about the symmetry of quantum phase space corresponding to the LCT group and its relation to the possible existence of sterile neutrinos. It is shown that the spin representation of the LCT group associated to the quantum phase, for a signature (1, 4), suggests the existence of sterile neutrinos and lead to a new way for describing them. The mathematical formalism developed in this work provides a new framework for the study of neutrinos physics.

## 1- Introduction

On one hand, the concept of Quantum Phase Space (QPS) has been introduced to extend the notion of phase space from classical physics to quantum physics [1-3]. On the other hand, sterile neutrinos are hypothetical particles that are expected to be important in the study and understanding of physics beyond the current standard model of particle physics. The existence of these particles are suggested both from theoretical and experimental sides [4-9]. The aim of this work is to give a description of the relations between sterile neutrinos and a symmetry group associated to QPS. This symmetry group is the group formed by multidimensional and relativistic Linear Canonical Transformations (LCTs) associated to a pentadimensional spacetime with signature (1, 4).

The concept of quantum phase space and its relation to LCTs have been introduced, developed and applied in the references [1-4]. The explicit relation between the existence of sterile neutrinos and a covariance principle associated to LCTs have been in particular firstly described in the reference [4]. The present work can be considered as a short review but some new ways of describing the sterile neutrinos states and their relation with the spin representation of LCTs are also developed here.

The notation system which is considered is based on the references [1-4]. Boldfaced letter like $\boldsymbol{p}$ are mainly used for quantum operator while normal letter like $p$ are used for eigenvalues.

## 2- Relativistic Quantum Phase Space

In classical physics, phase space is defined as the set of possible values of the momenta-coordinates pairs. It is not easy to extend this definition towards quantum physics because of the uncertainty relations which are consequences of the canonical commutation relations (CCRs). The uncertainty relations mean that there are limits to the precision with which the values of momenta-coordinates quantum observables pairs can be simultaneously known. It follows that the extension of the definition of phase space from classical to quantum physics is not trivial. The concept of relativistic quantum phase space gives a solution to this problem.

### 2.1 Relativistic generalization of the canonical commutation relations

The relativistic generalizations of the Canonical Commutation Relations (CCRs) are considered in the references [1-2], these relativistic CCRs are

$$\begin{cases} [\boldsymbol{p}_\mu, \boldsymbol{x}_\nu] = \boldsymbol{p}_\mu \boldsymbol{x}_\nu - \boldsymbol{x}_\nu \boldsymbol{p}_\mu = i\hbar \eta_{\mu\nu} \\ [\boldsymbol{p}_\mu, \boldsymbol{p}_\nu] = \boldsymbol{p}_\mu \boldsymbol{p}_\nu - \boldsymbol{p}_\nu \boldsymbol{p}_\mu = 0 \\ [\boldsymbol{x}_\mu, \boldsymbol{x}_\nu] = \boldsymbol{x}_\mu \boldsymbol{x}_\nu - \boldsymbol{x}_\nu \boldsymbol{x}_\mu = 0 \end{cases} \quad (1)$$

The relations (1) can be considered for a general theory corresponding to a pseudo-Euclidean space with signature $(N_+, N_-)$ and dimension $N = N_+ + N_-$. The $\eta_{\mu\nu}$ are then given by the following relation

$$\eta_{\mu\nu} = \begin{cases} 1 & if\ \mu = \nu = 0, 1, 2, \ldots, N_+ \\ -1 & if\ \mu = \nu = N_+ + 1, N_+ + 2, \ldots, N - 1 \\ 0 & if\ \mu \neq \nu \end{cases} \quad (2)$$

For the example of the pseudo-Euclidean space used in the current formulation of special relativity and relativistic quantum field theory, i.e. the Minkowski space, the signature is (1, 3) which means $N_+ = 1, N_- = 3, \eta_{00} = 1, \eta_{11} = \eta_{22} = \eta_{33} = -1$ and $\eta_{\mu\nu} = 0\ if\ \mu \neq \nu$.

## 2.2 Joint "momenta-coordinates" quantum states compatible with the CCRs

The introduction of the concept of Quantum Phase Space (QPS) needs the search for some kind of joint momenta-coordinates quantum states that are compatible with the CCRs and the uncertainty relations. It can be shown that the kind of states satisfying these criterion and which saturate the uncertainty relations are the states denoted $|\{\langle z_\mu \rangle\}\rangle = |\langle z \rangle\rangle$ corresponding to Gaussian-like wavefunctions [1-2]. The explicit expression of these wavefunctions are given in the following relation

$$\langle x^\mu | \{\langle z_\mu \rangle\} \rangle = \langle x | \langle z \rangle \rangle = \varphi_0(x) = \frac{e^{\frac{-\mathcal{B}_{\mu\nu}}{(\hbar)^2}(x^\mu - \langle x^\mu \rangle)(x^\nu - \langle x^\nu \rangle) - \frac{i}{\hbar}\langle p_\mu \rangle x^\mu + iK}}{[(2\pi)^N |det[\mathcal{X}_\nu^\mu]|]^{1/4}} \quad (3)$$

The parameters $\mathcal{B}_{\mu\nu}$ in the relation (3) are linked to the momenta-coordinates statistical variance-covariances, denoted $\mathcal{P}_{\mu\nu}, \mathcal{X}_{\mu\nu}$ and $\varrho_{\mu\nu}$, corresponding to the state $|\langle z \rangle\rangle$ itself and which are defined by the following relations

$$\begin{cases} \mathcal{P}_{\mu\nu} = \langle\langle z|(\boldsymbol{p_\mu} - \langle p_\mu \rangle)(\boldsymbol{p_\nu} - \langle p_\nu \rangle)|\langle z \rangle\rangle = \eta_{\mu\rho}\mathcal{P}_\nu^\rho \\ \mathcal{X}_{\mu\nu} = \langle\langle z|(\boldsymbol{x_\mu} - \langle x_\mu \rangle)(\boldsymbol{x_\nu} - \langle x_\nu \rangle)|\langle z \rangle\rangle = \eta_{\mu\rho}\mathcal{X}_\nu^\rho \\ \varrho_{\mu\nu} = \frac{1}{2}(\langle\langle z|(\boldsymbol{p_\mu} - \langle p_\mu \rangle)(\boldsymbol{x_\nu} - \langle x_\nu \rangle) + (\boldsymbol{x_\nu} - \langle x_\nu \rangle)(\boldsymbol{p_\mu} - \langle p_\mu \rangle)|\langle z \rangle\rangle) \end{cases} \quad (4)$$

with $\langle p_\mu \rangle = \langle\langle z|\boldsymbol{p_\mu}|\langle z \rangle\rangle$ and $\langle x_\mu \rangle = \langle\langle z|\boldsymbol{x_\mu}|\langle z \rangle\rangle$ the mean values of the momenta and coordinates operators associated to the state $|\langle z \rangle\rangle$ itself. The explicit expression of $\mathcal{B}_{\mu\nu}$ is [1-2]

$$\mathcal{B}_{\mu\nu} = \frac{(\hbar)^2}{4}\left(\eta_{\mu\rho} + \frac{2i}{\hbar}\varrho_{\mu\rho}\right)\tilde{\mathcal{X}}_\nu^\rho \quad (5)$$

in which the parameter $\tilde{\mathcal{X}}_\nu^\rho$ are related to $\mathcal{P}_{\mu\nu}, \mathcal{X}_{\mu\nu}$ and $\varrho_{\mu\nu}$ by the following relations

$$\begin{cases} \tilde{\mathcal{X}}_\mu^\rho \mathcal{X}_{\rho\nu} = \eta_{\mu\nu} \\ \mathcal{P}_{\mu\nu} = \frac{(\hbar)^2}{4}\tilde{\mathcal{X}}_{\mu\nu} + \varrho_{\mu\alpha}\tilde{\mathcal{X}}^{\alpha\beta}\varrho_{\nu\beta} \end{cases} \quad (6)$$

The term $|det[\mathcal{X}_\nu^\mu]|$ in the relation (3) is the absolute value of the determinant of the coordinate variances-covariances matrix with elements $\mathcal{X}_\nu^\mu = \eta^{\mu\rho}\mathcal{X}_{\rho\nu}$ and $K$ is a real number that doesn't depend on the coordinates $x^\mu$ ($e^{iK}$ is a unitary complex number).

The states $|\langle z \rangle\rangle$ defined by the relation (3) are called phase space states [1]. It can be shown that these states are also eigenstates of the operator $\boldsymbol{z_\mu}$ defined by the following relation [1-2]

$$\boldsymbol{z_\mu} = \boldsymbol{p_\mu} + \frac{2i}{\hbar}\mathcal{B}_{\mu\nu}\boldsymbol{x}^\nu \quad (7)$$

The corresponding eigenvalue equation is

$$z_\mu |\langle z \rangle\rangle = [\langle p_\mu \rangle + \frac{2i}{\hbar} \mathcal{B}_{\mu\nu} \langle x^\nu \rangle] |\langle z \rangle\rangle \quad (8)$$

with $\langle p_\mu \rangle = \langle\langle z | p_\mu | \langle z \rangle\rangle$ and $\langle x^\mu \rangle = \langle\langle z | x^\mu | \langle z \rangle\rangle$ the mean values of the operators $p_\mu$ and $x^\mu$ corresponding to the state $|\langle z \rangle\rangle$ itself. The momenta operators $p_\mu$ and their eigenstates $|p\rangle$ can be considered as asymptotic limits of the operators $z_\mu$ and their eigenstates $|\langle z \rangle\rangle$ when the parameters $\mathcal{B}_{\mu\nu}$ vanish. The states $|\langle z \rangle\rangle$ can also be considered as multidimensional and relativistic generalizations of what are called coherent states and squeezed states in the literature [10-11]. In fact, like these ones, the states $|\langle z \rangle\rangle$ correspond to Gaussian-like wavefunctions that saturate the uncertainty relations.

### 2.3 Definition of the quantum phase space

According to the reference [2], the quantum phase space can be defined as the set $\{\langle z_\mu \rangle\}$ of all possibles values of the mean values of the operators $z_\mu$ or equivalently as the set $\{(\langle p_\mu \rangle, \langle x_\mu \rangle)\}$ of the possibles values of the mean values of the momenta operators $p_\mu$ and coordinate operators $x_\mu$ for a given value of the $2N \times 2N$ momenta variance-covariance matrix $\begin{pmatrix} \mathcal{P} & \varrho \\ \varrho^T & \mathcal{X} \end{pmatrix}$

It follows from this definition that the structure of the quantum phase space depends explicitly on the values of the momenta-coordinates variance-covariances. It is this explicit dependence that make it compatible with the uncertainty principle. When all of the momenta-coordinates variance covariances are taken to be equal to zero (ignoring uncertainty principle), the quantum phase space as defined previously is reduced to a classical (non-quantum) phase space. It can also be remarked here that the momenta and coordinates variance-covariances can be possibly linked to thermodynamics parameters like temperature, pressure and volume shape and size as shown in the references [1, 3].

### 3- Symmetry group associated to a Quantum Phase Space

### 3.1 Linear Canonical Transformations (LCTs)

A natural symmetry group associated to a QPS is the group, denoted $\mathbb{T}$, formed by linear transformations mixing the momenta and coordinates operators and which leave covariant the CCRs [1-2]. These kind of transformations are the Linear Canonical Transformations (LCTs). Their explicit definition is given by the following relations

$$\begin{cases} p'_\mu = \mathbb{A}_\mu^\nu p_\nu + \mathbb{B}_\mu^\nu x_\nu \\ x'_\mu = \mathbb{C}_\mu^\nu p_\nu + \mathbb{D}_\mu^\nu x_\nu \\ [p'_\mu, x'_\nu] = [p_\mu, x_\nu] = i\hbar \eta_{\mu\nu} \\ [p'_\mu, p'_\nu] = [p_\mu, p_\nu] = 0 \\ [x'_\mu, x'_\nu] = [x_\mu, x_\nu] = 0 \end{cases} \quad (9)$$

It can be deduced from the relation (9) that the $N \times N$ matrices $\mathbb{a}, \mathbb{b}, \mathbb{c}, \mathbb{d}$ and $\eta$ corresponding to the parameters $\mathbb{a}_\mu^\nu, \mathbb{b}_\mu^\nu, \mathbb{c}_\mu^\nu, \mathbb{d}_\mu^\nu$ and $\eta_{\mu\nu}$ must satisfy the following relations [1-2]

$$\begin{cases} \mathbb{A}^T\eta\mathbb{D} - \mathbb{B}^T\eta\mathbb{C} = \eta \\ \mathbb{A}^T\eta\mathbb{B} - \mathbb{B}^T\eta\mathbb{A} = 0 \\ \mathbb{C}^T\eta\mathbb{D} - \mathbb{D}^T\eta\mathbb{C} = 0 \end{cases} \Leftrightarrow \begin{pmatrix} \mathbb{A} & \mathbb{C} \\ \mathbb{B} & \mathbb{D} \end{pmatrix}^T \begin{pmatrix} 0 & \eta \\ -\eta & 0 \end{pmatrix} \begin{pmatrix} \mathbb{A} & \mathbb{C} \\ \mathbb{B} & \mathbb{D} \end{pmatrix} = \begin{pmatrix} 0 & \eta \\ -\eta & 0 \end{pmatrix} \quad (10)$$

The relation (10) means that there is an isomorphism between the group $\mathbb{T}$ formed by the LCTs and the symplectic group $Sp(2N_+, 2N_-)$ [1-2].

According to the references [1-2], a LCT can be considered as corresponding to a change of frames of reference in the framework of relativistic quantum physics (change of observers). On the set of wavefuctions, LCTs are explicitly equivalent to integral transforms generalizing Fourier transform and fractional Fourier transforms as well known in signal processing and optics [12-15]. It is worth noting that the integral transformations corresponding to the LCTs leave covariant the form of the wavefunctions corresponding to the phase space states $|\langle z \rangle\rangle$ defined in the relation (3).

As shown in the reference [2], some other particular examples of LCTs are also the transformation corresponding to the elements of the orthogonal group $O(N_+, N_-)$. For the signature (1,3), this group corresponds to the Lorentz group $O(1,3)$. For the signature (1,4), one can have the de Sitter special relativity with the symmetry group $SO(1,4)$ [16]. The de Sitter special relativity can be associated to a de Sitter universe (vacuum with positive cosmological constant). As it will be shown, it is this signature (1, 4) that should be chosen to describe the link between quantum phase space symmetry and sterile neutrinos.

**3.2 Invariant quadratic operators and their eigenstates**

It can be shown that there are three kind of invariant quadratic operators that can be associated to the LCTs [1]

- A bosonic invariant quadratic operator denoted $\aleph$ given by the following relation

$$\aleph = \delta^{\mu\nu}\aleph_{\mu\nu} = \delta^{\mu\nu}\boldsymbol{z}_\mu^\dagger \boldsymbol{z}_\nu \quad (11)$$

with $\boldsymbol{z}_\mu$ and $\boldsymbol{z}_\mu^\dagger$ the operators defined by the following relations

$$\begin{cases} \boldsymbol{z}_\mu = \frac{1}{\sqrt{2}}(\boldsymbol{p}_\mu + i\boldsymbol{x}_\mu) = \frac{a_\mu^\nu}{\hbar}(z_\nu - \langle z_\nu \rangle) \\ \boldsymbol{z}_\mu^\dagger = \frac{1}{\sqrt{2}}(\boldsymbol{p}_\mu^\dagger - i\boldsymbol{x}_\mu^\dagger) = \frac{a_\mu^{\nu*}}{\hbar}(z_\nu^\dagger - \langle z_\nu \rangle^*) \end{cases} \quad [\boldsymbol{z}_\mu, \boldsymbol{z}_\nu^\dagger] = \delta_{\mu\nu} \quad (12)$$

with $\boldsymbol{p}_\mu$ and $\boldsymbol{x}_\mu$ the operators called « reduced momenta and reduced coordinates operators» and which are defined by the following relations [1-2]

$$\begin{cases} \boldsymbol{p}_\mu = \sqrt{2}[\frac{a_\mu^\nu}{\hbar}(\boldsymbol{p}_\nu - \langle p_\nu \rangle) - c_\mu^\nu(\boldsymbol{x}_\nu - \langle x_\nu \rangle)] \\ \boldsymbol{x}_\mu = \sqrt{2}\frac{\mathscr{b}_\mu^\nu}{\hbar}(\boldsymbol{x}_\nu - \langle x_\nu \rangle) \end{cases} \quad (13)$$

In the relations (12) and (13) : $\boldsymbol{p}_\mu$ and $\boldsymbol{x}_\mu$ are the momenta and coordinates operators, $z_\mu$ is the operator defined in the relation (4) and $a_\mu^\nu, \mathscr{b}_\mu^\nu$ and $c_\mu^\nu$ are respectively the elements of three matrices $a$, $b$ and $c$ which are related to the momenta-coordinates variance-covariance matrix $\begin{pmatrix} \mathcal{P} & \varrho \\ \varrho^T & \mathcal{X} \end{pmatrix}$ by the following relation

$$\begin{pmatrix} \mathcal{P} & \varrho \\ \varrho^T & \mathcal{X} \end{pmatrix} = \begin{pmatrix} b & 0 \\ 2acb & a \end{pmatrix}^T \begin{pmatrix} \eta & 0 \\ 0 & \eta \end{pmatrix} \begin{pmatrix} b & 0 \\ 2acb & a \end{pmatrix} \qquad (14)$$

According to the commutation relations in (12) and the relation (11), the operators $\pmb{z}_\mu$ and $\pmb{z}_\mu^\dagger$ have the properties of bosonic ladder operators and the operators $\aleph_{\mu\mu} = \pmb{z}_\mu^\dagger \pmb{z}_\mu$ have the properties of bosonic number operators. The common eigenstates of the operators $\aleph_{\mu\mu}$ and the operator $\aleph = \delta^{\mu\nu} \aleph_{\mu\nu}$, denoted $|n, \langle z \rangle\rangle$, can be deduced from the states $|\langle z \rangle\rangle$ defined in the relations (3) and (8) using the following relation

$$|n, \langle z \rangle\rangle = [\prod_{\mu=0}^{N-1} \frac{\left(\pmb{z}_\mu^\dagger\right)^{n_\mu}}{\sqrt{n_\mu!}}]|\langle z \rangle\rangle \qquad (15)$$

with $n$ referring to the set of $N$ nonnegative integers $n_0, n_1, \ldots, n_{N-1}$.

- A fermionic invariant quadratic operator, denoted $\pmb{\Sigma}$, that is associated with a spin representation of the LCTs [1]. Its expression is

$$\pmb{\Sigma} = \delta_{\mu\nu} \pmb{\Sigma}^{\mu\nu} = \delta_{\mu\nu} \pmb{\zeta}^{\mu\dagger} \pmb{\zeta}^\nu \qquad (16)$$

with $\pmb{\zeta}^{\mu\dagger}$ and $\pmb{\zeta}^\mu$ the operator defined by the following relations

$$\begin{cases} \pmb{\zeta}^\mu = \frac{1}{2}(\pmb{\alpha}^\mu + i\pmb{\beta}^\mu) \\ \pmb{\zeta}^{\mu\dagger} = \frac{1}{2}(\pmb{\alpha}^{\mu\dagger} - i\pmb{\beta}^{\mu\dagger}) \end{cases} \qquad (17)$$

in which $\pmb{\alpha}^\mu$ and $\pmb{\beta}^\mu$ are the generators of the Clifford algebra $\mathcal{Cl}(2N_+, 2N_-)$. One has the following anticommutation relations

$$\begin{cases} \pmb{\alpha}^\mu \pmb{\alpha}^\nu + \pmb{\alpha}^\nu \pmb{\alpha}^\mu = 2\eta^{\mu\nu} \\ \pmb{\beta}^\mu \pmb{\beta}^\nu + \pmb{\beta}^\nu \pmb{\beta}^\mu = 2\eta^{\mu\nu} \\ \pmb{\alpha}^\mu \pmb{\beta}^\nu + \pmb{\beta}^\nu \pmb{\alpha}^\mu = 0 \end{cases} \Leftrightarrow \begin{cases} \pmb{\zeta}^\mu \pmb{\zeta}^\nu + \pmb{\zeta}^\nu \pmb{\zeta}^\mu = 0 \\ \pmb{\zeta}^{\mu\dagger} \pmb{\zeta}^{\nu\dagger} + \pmb{\zeta}^{\nu\dagger} \pmb{\zeta}^{\mu\dagger} = 0 \\ \pmb{\zeta}^\mu \pmb{\zeta}^{\nu\dagger} + \pmb{\zeta}^{\nu\dagger} \pmb{\zeta}^\mu = \delta^{\mu\nu} \end{cases} \qquad (18)$$

According to the relations (16) and (17), the operators $\pmb{\zeta}^{\mu\dagger}$ and $\pmb{\zeta}^\mu$ have the properties of fermionic ladder operators and the operators $\pmb{\Sigma}^{\mu\mu} = \pmb{\zeta}^{\mu\dagger} \pmb{\zeta}^\mu$ has the properties of fermionic number operators.

- An hybrid or mixed (bosonic-fermionic) invariant quadric operator, denoted $(\mathbb{Z})^2$, which is given by the following relation

$$(\mathbb{Z})^2 = [\frac{1}{\sqrt{2}}(\pmb{\alpha}^\mu \pmb{p}_\mu + \pmb{\beta}^\mu \pmb{x}_\mu)]^2 = (\delta^{\mu\nu} \pmb{z}_\mu^\dagger \pmb{z}_\nu + \delta_{\mu\nu} \pmb{\zeta}^{\mu\dagger} \pmb{\zeta}^\nu) = \aleph + \pmb{\Sigma} \qquad (19)$$

The operator $\mathbb{Z}$ which is involved in the relation (19) is directly related to the concept of spin representation of LCTs described in [1]. It can be shown that the law of transformations of the reduced operator $\boldsymbol{p}_\mu$ and $\boldsymbol{x}_\mu$ defined in the relation (13) correspond to a pseudo-orthogonal representation of LCTs. These law of transformation can be put in the following form [1]

$$\begin{cases} \boldsymbol{p}'_\mu = \Pi_\mu^\nu \boldsymbol{p}_\nu + \Theta_\mu^\nu \boldsymbol{x}_\nu \\ \boldsymbol{x}'_\mu = -\Theta_\mu^\nu \boldsymbol{p}_\nu + \Pi_\mu^\nu \boldsymbol{x}_\nu \end{cases} \Leftrightarrow (\boldsymbol{p}' \quad \boldsymbol{x}') = (\boldsymbol{p} \quad \boldsymbol{x}) \begin{pmatrix} \Pi & -\Theta \\ \Theta & \Pi \end{pmatrix} \quad (20)$$

in which $\boldsymbol{p}$ and $\boldsymbol{x}$ are respectively the $1 \times N$ matrices corresponding to the $\boldsymbol{p}_\mu$ and $\boldsymbol{x}_\mu$. The $2N \times 2N$ matrix $\begin{pmatrix} \Pi & -\Theta \\ \Theta & \Pi \end{pmatrix}$ which corresponds to the LCT associated with the relation (20) is a pseudo-orthogonal matrix and belongs to the group $\mathbb{G}$ given by the following relation

$$\mathbb{G} \cong Sp(2N_+, 2N_-) \cap O(2N_+, 2N_-) \cong Sp(2N_+, 2N_-) \cap SO_0(2N_+, 2N_-) \quad (21)$$

in which $SO_0(2N_+, 2N_-)$ is the identity component of the indefinite special orthogonal group $SO(2N_+, 2N_-)$. The topological double cover of $\mathbb{G}$ is a subgroup $\mathbb{S}$ of the spin group $Spin(2N_+, 2N_-)$ which is the double cover of $SO_0(2N_+, 2N_-)$. The relation between $\mathbb{S}$ and $\mathbb{G}$ can be defined with a surjective covering map $u$ according to the relation

$$\begin{cases} u: \mathbb{S} \to \mathbb{G} \\ \mathcal{S} \mapsto \mathbb{g} = \begin{pmatrix} \Pi & -\Theta \\ \Theta & \Pi \end{pmatrix} \end{cases} \Leftrightarrow \begin{cases} (\boldsymbol{p}' \quad \boldsymbol{x}') = (\boldsymbol{p} \quad \boldsymbol{x}) \begin{pmatrix} \Pi & -\Theta \\ \Theta & \Pi \end{pmatrix} \\ \mathbb{Z}' = \mathcal{S}\mathbb{Z}\mathcal{S}^{-1} \end{cases} \quad (22)$$

with $\mathbb{Z}$ the operator which is involved in the relation (19) i.e. $(\mathbb{Z})^2$ is the hybrid invariant quadric operator.

The operators $\aleph_{\mu\mu} = \boldsymbol{z}_\mu^\dagger \boldsymbol{z}_\mu$, $\Sigma^{\mu\mu} = \zeta^{\mu\dagger}\zeta^\mu$ and the three invariant quadratic operators $\aleph = \delta_{\mu\nu}\Sigma^{\mu\nu}$, $\Sigma = \delta_{\mu\nu}\zeta^{\mu\dagger}\zeta^\nu$ and $(\mathbb{Z})^2 = \aleph + \Sigma$ have common eigenstates that are denoted $|n, \mathit{f}, \langle z \rangle\rangle$. The corresponding eigenvalue equations are [1]

$$\begin{cases} \aleph_{\mu\mu}|n, \mathit{f}, \langle z \rangle\rangle = \boldsymbol{z}_\mu^\dagger \boldsymbol{z}_\nu |n, \mathit{f}, \langle z \rangle\rangle = n_\mu |n, \mathit{f}, \langle z \rangle\rangle \\ \Sigma^{\mu\mu}|n, \mathit{f}, \langle z \rangle\rangle = \zeta^{\mu\dagger}\zeta^\mu |n, \mathit{f}, \langle z \rangle\rangle = \mathit{f}^\mu |n, \mathit{f}, \langle z \rangle\rangle \\ \aleph|n, \mathit{f}, \langle z \rangle\rangle = \delta^{\mu\nu}\boldsymbol{z}_\mu^\dagger \boldsymbol{z}_\nu |n, \mathit{f}, \langle z \rangle\rangle = |n||n, \mathit{f}, \langle z \rangle\rangle \\ \Sigma|n, \mathit{f}, \langle z \rangle\rangle = \delta_{\mu\nu}\zeta^{\mu\dagger}\zeta^\nu |n, \mathit{f}, \langle z \rangle\rangle = |\mathit{f}||n, \mathit{f}, \langle z \rangle\rangle \\ (\mathbb{Z})^2|n, \mathit{f}, \langle z \rangle\rangle = (\aleph + \Sigma)|n, \mathit{f}, \langle z \rangle\rangle = (|n| + |\mathit{f}|)|n, \mathit{f}, \langle z \rangle\rangle \end{cases} \quad (23)$$

In which $n$ is referring to the set of $N$ nonegative integers $n_0, n_1, \ldots n_{N-1}$ and $\mathit{f}$ is referring to the set of $N$ numbers $\mathit{f}^0, \mathit{f}^1, \ldots \mathit{f}^{N-1}$ that can only take the value 0 or 1 : $\mathit{f}^\mu = 0$ or $\mathit{f}^\mu = 1$. Then the eigenvalues $|n|$ and $|\mathit{f}|$ of $\aleph$ and $\Sigma$ are given by the following relations

$$\begin{cases} |n| = n_0 + n_1 + \cdots n_{N-1} \\ |\mathit{f}| = \mathit{f}^0 + \mathit{f}^1 \ldots + \mathit{f}^{N-1} \end{cases} \quad (24)$$

## 4- Relation between sterile neutrinos and quantum phase space symmetry

### 4.1 Sterile neutrinos

The existence of neutrinos was postulated by W. Pauli in 1930 to satisfy laws of conservations in beta decay. Three flavors of neutrinos were discovered successively: the electron neutrinos $\nu_e$ in 1956, the muon neutrino $\nu_\mu$ in 1962 and the tau neutrinos $\nu_\tau$ in 2000. Neutrinos have no electric charge. Within the Standard Model of particle physics, it is supposed that they are also massless. However the existence of neutrinos flavor oscillations implies that they have nonzero mass [2015 Nobel Prize in Physics]. Neutrinos mixing and oscillations are described with the Pontecorvo–Maki–Nakagawa–Sakata matrix (PMNS matrix) $[U]$. This matrix corresponds to expressions of neutrinos with definite flavor $\nu_e, \nu_\mu, \nu_\tau$ as superposition of neutrinos with definite mass denoted $\nu_1, \nu_2, \nu_3$

$$\begin{pmatrix} \nu_e \\ \nu_\mu \\ \nu_\tau \end{pmatrix} = \begin{pmatrix} U_e^1 & U_e^2 & U_e^3 \\ U_\mu^1 & U_\mu^2 & U_\mu^3 \\ U_\tau^1 & U_\tau^2 & U_\tau^3 \end{pmatrix} \begin{pmatrix} \nu_1 \\ \nu_2 \\ \nu_3 \end{pmatrix} \quad (25)$$

In the framework of the Standard Model, there are only left-handed neutrinos. Each neutrino field is paired with the corresponding charged leptons to have a $SU(2)_L$ doublet. The right-handed part of the charged leptons field are $U(1)_Y$ singlet [17-18]. Then, we have the following fields for the leptons sector of the Standard Model

$$\begin{pmatrix} \nu_{eL} \\ e_L \end{pmatrix}, \begin{pmatrix} \nu_{\mu L} \\ \mu_L \end{pmatrix}, \begin{pmatrix} \nu_{\tau L} \\ \tau_L \end{pmatrix}, e_R, \mu_R, \tau_R$$

Sterile neutrinos are hypothetical right-handed neutrinos. All of their Standard Model charges (weak isospin, weak hypercharge, electric charge and strong colors charges) are expected to be equal to zero. So they are "sterile" under electroweak and strong interactions. The existence of sterile neutrinos are suggested to solve some of the main problems related directly or indirectly to neutrinos: neutrinos masses, neutrinos oscillations anomalies (suggested in some experiments like LSND and MiniBooNE [8]), matter-antimatter asymmetry, dark matter, etc.

### 4.2 Description of sterile neutrinos from the spin representation of LCTs

The spin representation of the symmetry group $\mathbb{T} \cong Sp(2, 8)$ of a quantum phase space corresponding to the signature (1, 4) permits to deduce a classification of the elementary particles of the Standard Model with the prediction of the existence of sterile neutrinos [1-3]. The following operators can be considered for this purpose [1]

$$\begin{cases} \Sigma^{\mu\mu} = \zeta^{\mu\dagger}\zeta^\mu \\ I_3 = \frac{1}{2}(\Sigma^{00} + \Sigma^{44}) - \frac{1}{2} \\ Y_W = \Sigma^{00} - \frac{2}{3}(\Sigma^{11} + \Sigma^{22} + \Sigma^{33}) - \Sigma^{44} + 1 \\ Q = \Sigma^{00} - \frac{1}{3}(\Sigma^{11} + \Sigma^{22} + \Sigma^{33}) = I_3 + \frac{Y_W}{2} \end{cases} \quad (26)$$

According to the references [1-3], it can be explicitly checked that the eigenvalues of the operators $I_3$, $Y_W$ and $Q$ can be identified respectively with the weak isospin, weak hypercharge and electric charge of the leptons and quarks. Some of the possible values of these eigenvalues correspond to right-handed (sterile) neutrinos states and to their antiparticles. These states are common eigenstates of the operators $\Sigma^{\mu\mu} = \zeta^{\mu\dagger}\zeta^\mu$, $I_3, Y_W, Q, \Sigma, z^{\mu\dagger}z^\mu, \aleph$ and $(\mathbb{Z})^2$. Their explicit expressions are

$$\begin{cases} |n, \nu_R, \langle z\rangle\rangle = |n_0, n_1, n_2, n_3, n_4, 0,0,0,0,1, \langle z\rangle\rangle \\ |n, \bar{\nu}_R, \langle z\rangle\rangle = |n_0, n_1, n_2, n_3, n_4, 1,1,1,1,0, \langle z\rangle\rangle \end{cases} \quad (27)$$

The eigenvalues equations corresponding to the sterile neutrino states $|n, \nu_R, \langle z\rangle\rangle$ are given in the relation (28) below and those corresponding to antineutrinos are given in the relation (29). These eigenvalue equations are deduced from the general eigenvalues equations given in the relation (23) and on the relations (26) and (27)).

$$\begin{cases} \Sigma^{00}|n, \nu_R, \langle z\rangle\rangle = 0|n, \nu_R, \langle z\rangle\rangle \\ \Sigma^{11}|n, \nu_R, \langle z\rangle\rangle = 0|n, \nu_R, \langle z\rangle\rangle \\ \Sigma^{22}|n, \nu_R, \langle z\rangle\rangle = 0|n, \nu_R, \langle z\rangle\rangle \\ \Sigma^{33}|n, \nu_R, \langle z\rangle\rangle = 0|n, \nu_R, \langle z\rangle\rangle \\ \Sigma^{44}|n, \nu_R, \langle z\rangle\rangle = 1|n, \nu_R, \langle z\rangle\rangle \\ \Sigma|n, \nu_R, \langle z\rangle\rangle = 1|n, \nu_R, \langle z\rangle\rangle \\ I_3|n, \nu_R, \langle z\rangle\rangle = 0|n, \nu_R, \langle z\rangle\rangle \\ Y_W|n, \nu_R, \langle z\rangle\rangle = 0|n, \nu_R, \langle z\rangle\rangle \\ Q|n, \nu_R, \langle z\rangle\rangle = 0|n, \nu_R, \langle z\rangle\rangle \\ \aleph_{\mu\mu}|n, \nu_R, \langle z\rangle\rangle = n_\mu|n, \nu_R, \langle z\rangle\rangle \\ \aleph|n, \nu_R, \langle z\rangle\rangle = |n||n, \nu_R, \langle z\rangle\rangle \\ (\mathbb{Z})^2|n, \nu_R, \langle z\rangle\rangle = (|n|+1)|n, \nu_R, \langle z\rangle\rangle \end{cases} \quad (28)$$

$$\begin{cases} \Sigma^{00}|n, \bar{\nu}_R, \langle z\rangle\rangle = 1|n, \bar{\nu}_R, \langle z\rangle\rangle \\ \Sigma^{11}|n, \bar{\nu}_R, \langle z\rangle\rangle = 1|n, \bar{\nu}_R, \langle z\rangle\rangle \\ \Sigma^{22}|n, \bar{\nu}_R, \langle z\rangle\rangle = 1|n, \bar{\nu}_R, \langle z\rangle\rangle \\ \Sigma^{33}|n, \bar{\nu}_R, \langle z\rangle\rangle = 1|n, \nu_R, \langle z\rangle\rangle \\ \Sigma^{44}|n, \bar{\nu}_R, \langle z\rangle\rangle = 0|n, \bar{\nu}_R, \langle z\rangle\rangle \\ \Sigma|n, \bar{\nu}_R, \langle z\rangle\rangle = 4|n, \bar{\nu}_R, \langle z\rangle\rangle \\ I_3|n, \bar{\nu}_R, \langle z\rangle\rangle = 0|n, \bar{\nu}_R, \langle z\rangle\rangle \\ Y_W|n, \bar{\nu}_R, \langle z\rangle\rangle = 0|n, \bar{\nu}_R, \langle z\rangle\rangle \\ Q|n, \bar{\nu}_R, \langle z\rangle\rangle = 0|n, \bar{\nu}_R, \langle z\rangle\rangle \\ \aleph_{\mu\mu}|n, \bar{\nu}_R, \langle z\rangle\rangle = n_\mu|n, \bar{\nu}_R, \langle z\rangle\rangle \\ \aleph|n, \bar{\nu}_R, \langle z\rangle\rangle = |n||n, \bar{\nu}_R, \langle z\rangle\rangle \\ (\mathbb{Z})^2|n, \bar{\nu}_R, \langle z\rangle\rangle = (|n|+4)|n, \bar{\nu}_R, \langle z\rangle\rangle \end{cases} \quad (29)$$

The relations (26), (27), (28) and (29) lead to the prediction of the existence of a sterile neutrino, and a corresponding antiparticle, for each generations of fermions:

- A sterile neutrino state $|n, \nu_R, \langle z\rangle\rangle$ is a common eigenstate of the operators $\Sigma^{00}$ $\Sigma^{11}, \Sigma^{22}, \Sigma^{33}, \Sigma^{44}, \Sigma, I_3, Y_W$ and $Q$ with the eigenvalues $\not{f}^0 = 0, \not{f}^1 = 0, \not{f}^2 = 0, \not{f}^3 = 0, \not{f}^4 = 1, |\not{f}| = 1, I_3 = 0, Y_W = 0$ and $Q = 0$.

- A sterile antineutrino state $|n, \bar{\nu}_R, \langle z \rangle\rangle$ is a common eigenstate of the operators $\Sigma^{00}, \Sigma^{11}, \Sigma^{22}, \Sigma^{33}, \Sigma^{44}, \Sigma, I_3, Y_W$ and $Q$ with the eigenvalues $\not{f}^0 = 1, \not{f}^1 = 1, \not{f}^2 = 1, \not{f}^3 = 1, \not{f}^4 = 0, |\not{f}| = 4, I_3 = 0, Y_W = 0$ and $Q = 0$

If one consider a three generations of fermions, the relations (26), (27), (28) and (29) suggest the existence of three kind of steriles neutrinos (with their corresponding antiparticles). And according to these relation, the values of the weak isospin, weak hypercharge and electric charges depend only on the eigenvalues $\not{f}^\mu$ of the operators $\Sigma^{\mu\mu}$ and does not depend on the eigenvalues $n_0, n_1, n_2, n_3$ and $n_4$ of the operators $\aleph_{\mu\mu}$. It may be possible that these parameters $n_\mu$ may help in the understanding of the existence of multiple generations of fermions [3].

## 5- Conclusions

The spin representation of the symmetry group associated to a Relativistic Quantum Phase Space, for a signature (1, 4), suggests the existence of a right-handed sterile neutrino (and its antiparticle) for each flavor of neutrinos. The relevance of the signature (1, 4) may be related to the fact that it can be associated to a de Sitter space (vacuum with positive cosmological constant like the one that is considered in the current Standard, ΛCDM, Model of Cosmology [19-20]). It follows that the mathematical formalism associated to the concept of quantum phase space and its symmetry group may lead to new theoretical tools and framework for neutrinos physics, particle physics and cosmology. The existence of direct relations between the operators associated to the spin representation of the LCTs group and the weak isospin, weak hypercharges, electric charges and strong color charges as described in the relation (20) and in the references [1-3] suggest that there is a deep connection between phase space symmetry and particle properties. More in-depth studies on this connection could therefore lead to very interesting results and may help in the search for solutions to open questions in particle physics.

# REFERENCES

1. R. T. Ranaivoson, Raoelina Andriambololona, H. Rakotoson, R.H.M. Ravelonjato : Invariant quadratic operators associated with Linear Canonical Transformations and their eigenstates, arXiv: 2008.10602 [quant-ph], J. Phys. Commun. 6 095010, 2022.
2. R. T. Ranaivoson, Raoelina Andriambololona, H. Rakotoson, R. Raboanary: Linear Canonical Transformations in relativistic quantum physics, arXiv:1804.10053 [quant-ph], Physica Scripta , vol 96, No. 6., March 2021
3. R. H. M. Ravelonjato, R. T. Ranaivoson, Raoelina Andriambololona, R. Raboanary, H. Rakotoson, N. Rabesiranana: Quantum and Relativistic Corrections to Maxwell–Boltzmann Ideal Gas Model from a Quantum Phase Space Approach. Foundations of Physics 53:88 (2023)
4. Raoelina Andriambololona, R. T. Ranaivoson, H. Rakotoson, R. Raboanary: Sterile neutrino existence suggested from LCT covariance, arXiv:2109.03807 [hep-ph], *J. Phys. Commun.* 5 091001 (2021)
5. D. V. Naumov, Sterile Neutrino. A short introduction. arXiv:1901.00151 [hep-ph], EPJ Web of Conferences 207, 04004 (2019)
6. A.Boyarsky, M. Drewes, T.Lasserre,S.Mertens, O. Ruchayskiy, Sterile neutrino Dark Matter, Progress in Particle and Nuclear Physics, vol 104, January 2019
7. M. Drewes, The Phenomenology of Right Handed Neutrinos. arXiv: 1303.6912 [hep-ph], Int. J. Mod. Phys. E, 22, 1330019, (2013).
8. S. Böser, C. Buck, C. Giunti, J. Lesgourgues, L. Ludhova, S. Mertens, A. Schukraft , M. Wurm, Status of Light Sterile Neutrino Searches. arXiv: 1906.01739 [hep-ex], Prog. Part. Nucl. Phys., Vol. 111, (2020)
9. Janet M. Conrad, William C. Louis, Michael H. Shaevitz, The LSND and MiniBooNE Oscillation Searches at High Δm2, arXiv:1306.6494 [hep-ex], Annu. Rev. Nucl. Part. Sci. 63, 45 (2013)
10. T.G. Philbin : Generalized coherent states, Am. J. Phys. 82, 742 (2014)
11. B. Bagchi, R. Ghosh, A. Khare : A pedestrian introduction to coherent and squeezed states, Int. J. Mod. Phys. A35, 2030011 (2020)
12. K.B. Wolf, Integral Transforms in Science and Engineering. Plenum, New York (1979)
13. T.Z. Xu, B.Z. Li, Linear Canonical Transform and Its Applications. Science Press, Beijing, China (2013)
14. J. J. Healy, M. A. Kutay, H. M. Ozaktas , J. T. Sheridan, Linear Canonical Transforms: Theory and Applications. Springer, New York (2016)
15. T. Alieva, M.J. Bastiaans, Properties of the linear canonical integral transformation. J. Opt. Soc. Am. A/ Vol. 24, No. 11 (2007)
16. R. Aldrovandi, J. P. Beltran Almeida, J. G. Pereira, de Sitter special relativity, arXiv:gr-qc/0606122
17. J. Iliopoulos, Introduction to the Standard Model of the Electro-Weak Interactions, 2012 CERN Summer School of Particle Physics, Angers : France (2012), arXiv:1305.6779 [hep-ph] (2013)
18. P. Langacker, Introduction to the Standard Model and Electroweak Physics, arXiv:0901.0241[hep-ph] (2009)
19. A. Heavens : The cosmological model: an overview and an outlook, J. Phys.: Conf. Ser. (2008)
20. R. T. Génova-Santos : The establishment of the Standard Cosmological Model through observations, In: Kabáth, P., Jones, D., Skarka, M. (eds) Reviews in Frontiers of Modern Astrophysics. Springer, Cham. https://doi.org/10.1007/978-3-030-38509-5_11,arXiv:2001.08297[astro-ph.CO], (2020)